\documentclass[pra,reprint,showpacs]{revtex4-1}
\usepackage[T1]{fontenc}
\usepackage[latin9]{inputenc}
\usepackage{color}
\usepackage{amsthm}
\usepackage{amsmath}
\usepackage{amssymb}
\usepackage{graphicx}

\theoremstyle{plain}

\providecommand{\theoremname}{Theorem}

\begin{document}
\global\long\global\long\global\long\def\bra#1{\mbox{\ensuremath{\langle#1|}}}
\global\long\global\long\global\long\def\ket#1{\mbox{\ensuremath{|#1\rangle}}}
\global\long\global\long\global\long\def\bk#1#2{\mbox{\ensuremath{\ensuremath{\langle#1|#2\rangle}}}}
\global\long\global\long\global\long\def\kb#1#2{\mbox{\ensuremath{\ensuremath{\ensuremath{|#1\rangle\!\langle#2|}}}}}

\title{Classical simulation of fermionic linear optics augmented with noisy
ancillas}

\author{Micha\l{} Oszmaniec}

\email{oszmaniec@cft.edu.pl}

\author{Jan Gutt}

\author{Marek Ku\'{s}}

\address{Center for Theoretical Physics, Polish Academy of Sciences, Al. Lotników
32/46, 02-668 Warszawa}
\begin{abstract}
Fermionic linear optics is a model of quantum computation which is efficiently
simulable on a classical probabilistic computer. We study the problem of a
classical simulation of fermionic linear optics augmented with noisy auxiliary
states. If the auxiliary state can be expressed as a convex combination of pure
Fermionic Gaussian states, the corresponding computation scheme is classically
simulable. We present an analytic characterisation of the set of
convex-Gaussian states in the first non-trivial case, in which the Hilbert
space of the ancilla is a four-mode Fock space. We use our result to solve an
open problem recently posed by De Melo et al. \cite{powernoisy2013} and to
study in detail the geometrical properties of the set of convex-Gaussian
states.
\end{abstract}


\maketitle
For any model of quantum computation it is vital to characterise its
computational power. Probably the most important practical question
is how a given model compares to universal classical, respectively
quantum, computation. If protocols allowed by the model are efficiently
simulable on a classical computer, the corresponding physical system
may be accessible to numerical studies, but is unlikely to be a suitable
candidate for a quantum computer. On the other hand, simulability
by quantum circuits ensures that the underlying physics can be effectively
studied using a quantum computer \cite{Feynman1982}. Lastly, if the
resources provided by the model enable one to implement a universal
quantum computation, the corresponding physical system is a candidate
for the realization of a quantum computer.

Many physically motivated models of quantum computation are defined by
specifying the available set of initial states, gates and measurements
\cite{Gottesman1998,Briegel2003}. The Fermionic Linear Optics (FLO) model of
quantum computation, introduced in \cite{Terhal2002}, is the fermionic analogue
of well-known bosonic linear optics \cite{Kok2007,Aaronson2011}. In the FLO
model the class of allowed operations includes: preparation of the vacuum
state, free unitary fermionic evolution, and occupation number measurements.
Fermionic linear optics describes systems of non-interacting fermions, i.e.
fermionic systems that can be described exactly by the Bogolyubov mean field
theory. The model of computation based on FLO alone is not computationally
universal and can be effectively simulated by a classical probabilistic
computer \cite{Terhal2002,BravyiKoeningSimul}. Nevertheless, the physics beyond
FLO is rich and captures a number of systems of interest for condensed matter
physics, including Kitaev's Majorana chain \cite{Kitaev2001}, honeycomb model
\cite{Kitaev2006}, $\nu=5/2$ fractional Quantum Hall systems \cite{Alicea2012}.
These systems possess a topological order and can be used as fault-tolerant
quantum memories \cite{Mazza2013} or for the Topological Quantum Computation
(TQC) with Ising anyons \cite{Kitaev2006}. This motivates an interest in
extending FLO in such a way that the resulting model will become
computationally universal. In the present paper we study the computational
power of the model in which FLO is augmented with a noisy ancilla state. In
this model, introduced in \cite{powernoisy2013}, the traditional scheme of FLO
is extended by introducing additional fermionic modes in which, at the
beginning of computation, one stores certain number of copies of a, perhaps
noisy, ancilla state. Initial state of the system in the original
``computational'' modes remains the vacuum state and the class of allowed
operations remains intact. This model of computation is analogous to other
ancilla-assisted models of quantum computation (such as ancilla-assisted
computation with Clifford gates \cite{Bravyi2005,Veitch2014} or
ancilla-assisted topological quantum computing with Ising anyons
\cite{universalfracBravyi}). Depending on the properties of the auxiliary state
it may be possible to implement, with the help of traditional FLO operations,
gates that are necessary for computational universality on registers describing
the actual computation. In \cite{powernoisy2013} the authors showed that
whenever the state of the ancilla is convex-Gaussian, i.e. can be written as a
convex combination of projectors onto pure Gaussian states (see definition
bellow), the corresponding model of computation remains classically simulable.
On the other hand, some ancilla states \cite{Bravyi2005} do promote the model
to be computationally universal.

In the current paper we present a complete analytical characterisation of the
set of convex-Gaussian states for a special case when the Fock space of the
ancilla has four modes. This is the lowest-dimensional non-trivial case, as for
two and three modes all even pure states are Gaussian \cite{powernoisy2013}.
Moreover, the four mode Fock space is of special relevance as some states from
this space can be used to make FLO and TQC with Ising anyons computationally
universal \cite{universalfracBravyi}. A particular example is a non-Gaussian
pure state $\ket{a_{8}}$ (see definition below). We use our analytical
criterion to show that for a noise strength $p\geq p_{cr}=\frac{8}{11}$, the
depolarised state $\ket{a_{8}}$ becomes convex-Gaussian and thus useless for
computational purposes. We thus settle the open problem posed in a recent paper
by de Melo et. al. \cite{powernoisy2013}. In addition to the exact analytical
criterion we explore, with the use of group-theoretical methods and tools of
entanglement theory, the geometry of the set of convex-Gaussian states in the
space of all density operators in a four mode Fock space.

We first introduce the necessary notation and briefly describe the
FLO model of computation and its classical simulation. The Hilbert
space describing a fermionic system whose particles can be in $d$
modes is a Fock space which we denote by $\mathrm{Fock}\left(\mathbb{C}^{d}\right)$.
On this space we have the action of the standard creation and annihilation
operators: $a_{i}^{\dagger}$, $a_{i}$ , $i=1,\ldots,d$, which satisfy
canonical anti-commutation relations. The whole Fock space is spanned
by the set of orthonormal Fock states:
\begin{equation}
\ket{n_{1},\ldots,n_{d}}=\left(a_{1}^{\dagger}\right)^{n_{1}}\cdots\left(a_{d}^{\dagger}\right)^{n_{d}}\ket 0\,,\,\label{eq:fock states}
\end{equation}
where $\ket 0$ is the Fock vacuum. It what follows we will consider
only even operators, i.e., operators commuting with the total parity
operator $Q=\prod_{k=1}^{d}\left(\mathbb{I}-2a_{k}^{\dagger}a_{k}\right)$.
The operator $Q$ has eigenvalues $\pm1$. The Fock space decomposes
onto eigenspaces of $Q$, $\mathrm{Fock}\left(\mathbb{C}^{d}\right)=\mathrm{Fock}_{+}\left(\mathbb{C}^{d}\right)\oplus\mathrm{Fock}_{-}\left(\mathbb{C}^{d}\right)$,
where $\mathrm{Fock}_{+}\left(\mathbb{C}^{d}\right)$ and $\mathrm{Fock}_{-}\left(\mathbb{C}^{d}\right)$
are spanned by Fock states with respectively even and odd number of
excitations. The corresponding orthogonal projectors onto these subspaces
are $\mathbb{P}_{\pm}=\frac{1}{2}\left(\mathbb{I}\pm Q\right)$. It
is convenient to introduce Majorana fermion operators \cite{Bravyi2004,powernoisy2013}:
$c_{2k-1}=a_{k}+a_{k}^{\dagger}$, $c_{2k}=i\left(a_{k}-a_{k}^{\dagger}\right)$
, $k=1,\ldots,d$. One checks that they are Hermitian and satisfy
anti-commutation relations $\left\{ c_{k},\, c_{l}\right\} =2\delta_{kl}$.
The operator $Q$ takes the from $Q=i^{d}\prod_{k=1}^{2d}c_{k}$.
An Hermitian operator $X$ is even if and only if it can be written
as a polynomial in Majorana operators involving only monomials of
even degree,
\begin{equation}
X=\alpha_{0}\mathbb{I}+\sum_{k=1}^{d}i^{k}\sum_{1\leq l_{1}\leq l_{2}\leq\ldots\leq l_{2k}\leq2d}\alpha_{l_{1}l_{2}\ldots l_{2k}}c_{l_{1}}c_{l_{2}}\ldots c_{l_{2k}}\,,\label{eq:decomposition}
\end{equation}
where all the coefficients $\alpha_{0}$ and $\alpha_{l_{1}l_{2}\ldots l_{2k}}$
are real. For an even mixed state $\rho$ the correlation matrix $M$ is defined
by $M_{ij}=\frac{i}{2}\mathrm{Tr}\left(\rho\,\left[c_{k},\,
c_{l}\right]\right)\,,\, k,l=1,\ldots,2d\,;$ it is real and antisymmetric. Pure
fermionic Gaussian states, which we denote by $\mathcal{G}$, are by definition
\cite{powernoisy2013}, states for which the correlation matrix is orthogonal,
\begin{equation}
\mathcal{G}=\left\{ \ket{\psi}\in\mathcal{H}_{\mathrm{Fock}}\left(\mathbb{C}^{N}\right)\,|\, MM^{T}=\mathbb{I}_{2N}\right\} .\,\label{eq:gauss}
\end{equation}
Pure fermionic Gaussian states, by virtue of the fermionic Wick theorem
\cite{Bravyi2004}, are fully determined by their correlation matrix
$M$. A given mixed state $\rho$ is called convex-Gaussian if and
only if it can be expressed as a convex combination of pure Gaussian
states, $\rho=\sum_{i}p_{i}\kb{\psi_{i}}{\psi_{i}}\,,$ where $\sum_{i}p_{i}=1$,
and for each $\ket{\psi_{i}}\in\mathcal{G}$ is a pure Gaussian state.
We denote the set of convex-Gaussian %
\footnote{It is important to underline the difference between convex-Gaussian
states and general Gaussian states. The latter class consists of states
of the form $\rho=K\cdot\mathrm{exp}\left(\sum_{i\neq j}h_{ij}c_{i}c_{j}\right)$,
where $K$ is a normalisation constant. Every Gaussian state is convex-Gaussian,
but the converse is not true. %
} states by $\mathcal{G}^{c}$ expressing the fact that this set is
the convex hull of $\mathcal{G}$ in the space of Hermitian operators
on $\mathrm{Fock}\left(\mathbb{C}^{N}\right)$. Note that we have
an analogy with the separability problem, where the class of separable
mixed states is defined as the convex hull of the set of pure product
states.

We now briefly recall the computation model based on fermionic linear optics
introduced in \cite{powernoisy2013}. The allowed operations in the model are:
(i) preparation of the Fock vacuum $\ket 0$, (ii) measurement of the occupation
numbers
$a_{k}^{\dagger}a_{k}=\frac{1}{2}\left(\mathbb{I}+ic_{2k-1}c_{2k}\right)$ for
any mode $k$ , (iii) evolution under the von Neumann equation,
$\frac{d}{dt}\rho=-i\left[H,\,\rho\right]$, for time $t$. The Hamiltonian
$H=i\sum_{k,l=1}^{2N}h_{kl}c_{k}c_{l}$ is an arbitrary Hamiltonian quadratic in
Majorana operators. Operations (i), (ii) and (iii) can be preformed in
arbitrary order and may depend upon measurement results obtained during
previous stages of the computation. The protocol concludes with the final
measurement whose (binary) outcome is the result of the computation. The above
model of computation can be efficiently simulated in polynomial time on a
probabilistic classical computer (a classical computer having access to random
bits). The proof relies on the fact that the state remains Gaussian along the
computation and that there exist update rules for the correlation matrix $M$
that have low computational complexity. We now extend the above model by
allowing (iv) multiple usage of the ancilla state $\rho$ that is stored in
auxiliary $m$ modes ($\rho$ is a state on
$\mathrm{Fock}\left(\mathbb{C}^{m}\right)$). If $k$ auxiliary states are
available, the total Hilbert space of the system becomes
$\mathrm{Fock}\left(\mathbb{C}^{d}\right)\otimes\left(\mathrm{Fock}\left(\mathbb{C}^{m}\right)\right)^{\otimes
k}=\mathrm{Fock}\left(\mathbb{C}^{d+k\cdot m}\right)$ and we allow arbitrary
operations of the form (ii) and (iii) to be performed on the initial state of
the total system of the form $\kb 00\otimes\rho^{\otimes k}$. The computation
model (i-iv) can be effectively classically simulated if the auxiliary state
$\rho$ is convex-Gaussian i.e. $\rho=\sum_{i}p_{i}\kb{\psi_{i}}{\psi_{i}}$ for
$\ket{\psi_{i}}\in\mathcal{G}$ \cite{powernoisy2013}. This is the reason why
the characterisation of the convex-Gaussian states $\mathcal{G}^{c}$ is
important for this model of computation. The simulation scheme consists of
sampling pure Gaussian states $\left\{ \ket{\psi_{i}}\right\} $ according to
the probability distribution $\left\{ p_{i}\right\} $ followed by the classical
simulation of the of the evolution of Gaussian states described in
\cite{Terhal2002,Bravyi2004}.  We would like to point out that convex-Gaussian
ancilla states lead to an effectively classically simulable model also when one
replaces FLO with its dissipative counterpart, recently introduced in
\cite{BravyiKoeningSimul}. For this reason results presented in this work are
also valid for the dissipative FLO.

We can now give a complete analytical characterisation of the set
of convex-Gaussian states for the special case of $\mathrm{Fock}\left(\mathbb{C}^{4}\right)$.
Let $\rho$ be an arbitrary even mixed state on $\mathrm{Fock}\left(\mathbb{C}^{4}\right)$
having the decomposition (\ref{eq:decomposition}). Let $\rho_{+}=\mathbb{P}_{+}\rho\mathbb{P}_{+}$,
$\rho_{-}=\mathbb{P}_{-}\rho\mathbb{P}_{-}$ denote restrictions of
$\rho$ to $\mathrm{Fock}_{+}\left(\mathbb{C}^{4}\right)$ and $\mathrm{Fock}_{-}\left(\mathbb{C}^{4}\right)$,
respectively. By $\tilde{X}$ we denote the ``complex conjugate''
of the operator $X$, i.e., an operator constructed from $X$ by changing
all $i$ to $-i$ in the decomposition (\ref{eq:decomposition}).
Let us introduce non-negative functions $C_{+}$ and $C_{-}$ that
are the analogues of Ulhmann-Wooters concurrence describing entanglement
in two qubit systems \cite{Uhlmann1999}. They are defined by
\begin{equation}
C_{\pm}\left(\rho\right)=\mathrm{max}\left\{ 0,\,\lambda_{1}^{\pm}-\sum_{k=2}^{8}\lambda_{k}^{\pm}\right\} ,\label{eq:gen concurrence}
\end{equation}
where
$\left(\lambda_{1}^{\pm},\,\lambda_{2}^{\pm},\ldots,\,\lambda_{8}^{\pm}\right)$
denote non-increasingly ordered eigenvalues of the operator
$\sqrt{\rho_{\pm}\tilde{\rho}_{\pm}}$. Convex-Gaussianity of $\rho$ is
characterised by the values of these generalised concurrences,
\begin{equation}
\rho\,\text{ is convex-Gaussian }\Longleftrightarrow C_{+}\left(\rho\right)=C_{-}\left(\rho\right)=0\,.\label{eq:criterion}
\end{equation}
Application of methods from entanglement theory enable us to give
a detailed description of the geometry of convex-Gaussian states in
$\mathrm{Fock}_{+}\left(\mathbb{C}^{4}\right)$ %
\footnote{Analogous results can be easily derived for convex-Gaussian states
in $\mathrm{Fock}_{-}\left(\mathbb{C}^{4}\right)$ and general even
states $\rho$ in $\mathrm{Fock}\left(\mathbb{C}^{4}\right)$, but
we do not present them here for simplicity.%
}. For a given state $\rho,$ supported in $\mathrm{Fock}_{+}\left(\mathbb{C}^{4}\right)$,
we provide a measure of its distance to the set of convex-Gaussian
states $\mathcal{G}^{c}$ in terms of Uhlmann fidelity with respect
to the set $\mathcal{G}^{c}$. The following formula holds
\begin{equation}
F_{\mathrm{Gauss}}\left(\rho\right)=\mathrm{max}_{\sigma\in\mathcal{G}^{c}}F\left(\rho,\sigma\right)=\frac{1}{2}+\frac{1}{2}\sqrt{1-C_{+}^{2}\left(\rho\right)}\,,\label{eq:Fidelity}
\end{equation}
where $F\left(\rho,\sigma\right)=\left(\mathrm{tr}\left[\sqrt{\sqrt{\rho}\sigma\sqrt{\rho}}\right]\right)^{2}$
denotes Uhlmann fidelity between states $\rho$ and $\sigma$. Using
Fuchs-van de Graaf inequalities \cite{Fuchs1999} we bound the statistical
(trace) distance \cite{Bengtsson2006} of any state $\rho$ on $\mathrm{Fock}_{+}\left(\mathbb{C}^{4}\right)$
to the set of convex-Gaussian states, $D\left(\rho,\mathcal{G}^{c}\right)=\mathrm{min}_{\sigma\in\mathcal{G}^{c}}\frac{1}{2}\sqrt{\mathrm{tr}\left(\left(\rho-\sigma\right)^{2}\right)}$,
by
\begin{equation}
1-\sqrt{F_{\mathrm{Gauss}}\left(\rho\right)}\leq D\left(\rho,\mathcal{G}^{conv}\right)\leq\sqrt{1-F_{\mathrm{Gauss}}\left(\rho\right)}\,.\label{eq:statistical distance}
\end{equation}
Inequalities (\ref{eq:statistical distance}) together with (\ref{eq:Fidelity})
show that for a non convex-Gaussian state $\rho$ supported in $\mathrm{Fock}_{+}\left(\mathbb{C}^{4}\right)$
the generalised concurrence $C_{+}\left(\rho\right)$ can be used
to assess the resilience of the property of being non convex-Gaussian
against noise.

Before proceeding to the proofs of (\ref{eq:criterion}) and (\ref{eq:Fidelity}),
we use (\ref{eq:criterion}) to give the noise threshold $p_{cr}$
above which a depolarisation of the state $\ket{a_{8}}\in\mathrm{Fock}\left(\mathbb{C}^{4}\right)$
becomes convex-Gaussian. In other words we consider a state
\begin{equation}
\rho\left(p\right)=\left(1-p\right)\kb{a_{8}}{a_{8}}+p\frac{\mathbb{I}}{16}\,,\label{eq:depolarised a8}
\end{equation}
where $p\in[0,\,1${]}, $\mathbb{I}$ is the identity operator, and
\[
\kb{a_{8}}{a_{8}}=\frac{1}{16}\left(\mathbb{I}+S_{1}\right)\left(\mathbb{I}+S_{2}\right)\left(\mathbb{I}+S_{3}\right)\left(\mathbb{I}+Q\right)
\]
is a pure state which can be used to implement a $\mathrm{CNOT}$ gate that is
needed to promote FLO to be computationally universal
\cite{universalfracBravyi,powernoisy2013}. The problem of finding $p_{cr}$ was
considered in \cite{Terhal2002} where authors showed that $\rho\left(p\right)$
is non convex-Gaussian for $p\leq\frac{8}{15}$ and is convex-Gaussian for
$p\geq\frac{8}{9}$. Application of (\ref{eq:criterion}) to (\ref{eq:depolarised
a8}) is straightforward because
$\ket{a_{8}}\in\mathrm{Fock}\left(\mathbb{C}^{4}\right)$ and
$\tilde{\rho}\left(p\right)=\rho\left(p\right)$. Simple algebra shows that
$\rho(p)$ is convex-Gaussian if and only if $p\geq\frac{8}{11}=p_{cr}$. This
result is particularly interesting as it opens a possibility for existence of
more noise-resilient protocols of distillation of the state pure $\ket{a_{8}}$
from copies of a noisy state $\rho\left(p\right)$ via FLO or TQC with Ising
Anyons (the protocol based on TQC introduced in \cite{universalfracBravyi}
works for $p\leq0.4$).

We now prove our criterion (\ref{eq:criterion}). Let us first note that pure
fermionic Gaussian states have a fixed parity. In other words:
$\mathcal{G}=\mathcal{G}_{+}\cup\mathcal{G}_{-}$, where
$\mathcal{G}_{\pm}\subset\mathrm{Fock}_{\pm}\left(\mathbb{C}^{m}\right)$. For
this reason it is enough to consider the problem of convex-Gaussianity
separately on $\mathrm{Fock}_{\pm}\left(\mathbb{C}^{m}\right)$. In other words
an even state $\rho$ is convex-Gaussian if and only if both $\rho_{+}$ and
$\rho_{-}$ are convex-Gaussian. We show below that there exist antiunitary
operators $\theta_{\pm}$, each acting on
$\mathrm{Fock}_{\pm}\left(\mathbb{C}^{4}\right)$, such that
\begin{equation}
\ket{\psi}\in\mathcal{G}_{\pm}\Longleftrightarrow C_{\pm}\left(\ket{\psi}\right)=\left|\bk{\psi}{\theta_{\pm}\psi}\right|=0\,.\label{eq:pure concurrence}
\end{equation}
We can now use the Uhlmann-Wooters construction \cite{Uhlmann1999}
to compute the convex roof extension of $C_{\pm}$, $C_{\pm}\left(\sigma\right)=\mathrm{inf}_{\sum\kb{\psi_{i}}{\psi_{i}}=\sigma}\left(\sum_{i}C_{\pm}\left(\ket{\psi_{i}}\right)\right)$,
for $\sigma$ a non-negative operator on $\mathrm{Fock}_{\pm}\left(\mathbb{C}^{m}\right)$.
From the definition of the convex roof extension and the discussion
above we have that $\rho$ is convex-Gaussian if and only if $C_{+}\left(\rho_{+}\right)=C_{-}\left(\rho_{-}\right)=0$.
Explicit formulas for $C_{\pm}\left(\rho_{\pm}\right)$ are given
by (\ref{eq:gen concurrence}), where $\left(\lambda_{1}^{\pm},\,\lambda_{2}^{\pm},\ldots,\,\lambda_{8}^{\pm}\right)$
denote non-increasingly ordered eigenvalues of the operator $\sqrt{\rho_{\pm}\theta_{\pm}\rho_{\pm}\theta_{\pm}}$
\cite{Uhlmann1999} . From the details of the Uhlmann-Wooters construction
it follows that for convex-Gaussian states supported in $\mathrm{Fock}_{\pm}\left(\mathbb{C}^{4}\right)$
we need at most $\mathcal{N}=8$ pure Gaussian states in the convex
decomposition (this is a consequence of the fact that Hadamard matrices
exist in dimension $8$ which is the dimension of $\mathrm{Fock}_{+}\left(\mathbb{C}^{4}\right)$).
Consequently, for an arbitrary convex-Gaussian state in $\mathrm{Fock}\left(\mathbb{C}^{4}\right)$
this number equals $\mathcal{N}=16$, much smaller than the upper
bound $\tilde{\mathcal{N}}=48$ obtained in \cite{powernoisy2013}.
The existence of antiunitary operators $\theta_{\pm}$ follows from
group-theoretical interpretation of pure Gaussian states $\mathcal{G}_{\pm}$.
The group of Bogolyubov transformations is precisely $\mathrm{Spin}\left(2m\right),$
a compact semi-simple Lie group. The Hilbert space $\mathrm{Fock}\left(\mathbb{C}^{m}\right)$
decomposes into two irreducible representations of $\mathrm{Spin}\left(2m\right)$:
$\mathrm{Fock}_{+}\left(\mathbb{C}^{m}\right)$ and $\mathrm{Fock}_{-}\left(\mathbb{C}^{m}\right)$
respectively. Sets of pure Gaussian states $\mathcal{G}_{\pm}$ are
precisely the ``highest weight'' orbits of this group in $\mathrm{Fock}_{\pm}\left(\mathbb{C}^{m}\right)$
\cite{Manivel2009}. Semi-simple compact Lie groups $K$ and irreducible
representations $\mathcal{H}^{\lambda}$ %
\footnote{Irreducible representations of semi-simple Lie groups $K$ are characterised
by a so-called highest weight $\lambda$. The highest weight $\lambda$
is a generalisation of the ``total spin'' irreducible representations
of $SU(2)$. For a comprehensive introduction to the representation
theory of semi-simple Lie groups see \cite{Hall2003}.%
} admitting an antiunitary operator $\theta:\mathcal{H}^{\lambda}\rightarrow\mathcal{H}^{\lambda}$
detecting the orbit through the highest-weight vector had been classified
in \cite{detection2012}. In order to guarantee the existence of such
$\theta$ it suffices to check that the following decomposition holds:
\begin{equation}
\mathcal{H}^{\lambda}\vee\mathcal{H}^{\lambda}=\mathcal{H}^{2\lambda}\oplus\mathcal{H}_{0}\,,\label{eq:irrep decomp}
\end{equation}
where $\vee$ denotes the symmetric tensor product of Hilbert spaces
(thus, $\mathcal{H}^{\lambda}\vee\mathcal{H}^{\lambda}$ becomes a
representation of $K$), $\mathcal{H}^{2\lambda}$ is an irreducible
representation of $K$ characterised by the highest weight $2\lambda$,
and $\mathcal{H}_{0}$ is a trivial (one dimensional) representation
of $K$. From the construction of an antiunitary $\theta$ presented
in \cite{detection2012} it follows that it is $K$ invariant: $k\theta k^{\dagger}=\theta$,
for all elements $k$ of the Lie group $K$. In our case we have $K=\mathrm{Spin}\left(8\right)$
and $\mathcal{H}^{\lambda}=\mathrm{Fock}_{+}\left(\mathbb{C}^{4}\right)$
or $\mathcal{H}^{\lambda}=\mathrm{Fock}_{-}\left(\mathbb{C}^{4}\right)$.
For these particular representations the decomposition (\ref{eq:irrep decomp})
indeed holds (see for example \cite{Manivel2009}) and thus existence
of $\mathrm{Spin}\left(8\right)$-invariant antiunitaries $\theta_{\pm}$
is guaranteed. We conclude the proof of (\ref{eq:criterion}) showing
that $\theta_{\pm}X\theta_{\pm}=\tilde{X}$ %
\footnote{In fact one can combine antiunitary operators $\theta_{\pm}$ to get
an antiunitary operator $\tilde{\theta}$ acting on the total Fock
space and having similar properties. For simplicity we do not present
this construction here. %
} for every operator $X$ supported in either $\mathrm{Fock}_{+}\left(\mathbb{C}^{4}\right)$
or $\mathrm{Fock}_{-}\left(\mathbb{C}^{4}\right)$. We present here
a proof only for the even case. The desired property of $\theta_{+}$
follows from its invariance under the action of $\mathrm{Spin}\left(8\right)$.
Since $\mathrm{Spin}\left(8\right)$ is generated by anti-Hermitian
operators $c_{i}c_{j}$, it follows that $c_{i}c_{j}\theta_{+}=\theta_{+}c_{i}c_{j}$
for every pair of Majorana operators. Using the fact that $\theta_{+}$
satisfies (as every antiunitary operator) $\theta_{+}^{2}=\mathbb{I}$,
$\theta i=-i\theta$ and noting that every operator $X$ with support
in $\mathrm{Fock}_{+}\left(\mathbb{C}^{4}\right)$ is an even operator
(and thus has a decomposition (\ref{eq:decomposition})) proves $\theta_{\pm}X\theta_{\pm}=\tilde{X}$.
This concludes the proof of (\ref{eq:criterion}).

We can now describe the action of $\theta_{+}$ on pure states: expressing
$c_{i}c_{j}$ in terms of creation and annihilation operators, and
using antilinearity of $\theta_{+}$, one checks that $a_{k}a_{l}\theta_{+}=\theta_{+}a_{k}^{\dagger}a_{l}^{\dagger}$,
together with its conjugate $a_{k}^{\dagger}a_{l}^{\dagger}\theta_{+}=\theta_{+}a_{k}a_{l}$.
Furthermore, we have $a_{k}^{\dagger}a_{l}^{\dagger}\theta_{+}\ket 0=\theta_{+}a_{k}a_{l}\ket 0=0$
so that $\theta_{+}\ket 0$ is a maximally occupied state. We can
fix the phase ambiguity in $\theta_{+}$ to ensure $\theta_{+}\ket 0=$$\ket{1,1,1,1}$.
It then follows that $\theta_{+}$ sends a state $\ket{n_{1},n_{2},n_{3},n_{4}}$
in $\mathrm{Fock}_{+}\left(\mathbb{C}^{4}\right)$ to $(-1)^{N/2}\ket{\bar{n}_{1},\bar{n}_{2},\bar{n}_{3},\bar{n}_{4}}$,
where $\bar{0}=1$, $\bar{1}=0$ and $N=n_{1}+n_{2}+n_{3}+n_{4}.$
An analogous formula holds for the action of $\theta_{-}$ on $\mathrm{Fock}_{-}\left(\mathbb{C}^{4}\right)$,
so that we can write universally $\theta_{\pm}$$\ket{n_{1},n_{2},n_{3},n_{4}}$=$(-1)^{\lfloor N/2\rfloor}\ket{\bar{n}_{1},\bar{n}_{2},\bar{n}_{3},\bar{n}_{4}}$,
thus allowing one to view $\theta_{\pm}$ as implementing an excitation-hole
duality.

In order to verify (\ref{eq:Fidelity}) we describe the geometry of
the action of $\mathrm{Spin\left(8\right)}$ on pure states in $\mathrm{Fock}_{+}\left(\mathbb{C}^{4}\right)$.
We show that the concurrence of a pure state, $C_{+}(\ket{\psi})=|\bk{\psi}{\theta_{+}\psi}|$,
parametrises orbits of $\mathrm{Spin}\left(8\right)$ in $\mathrm{Fock}_{+}\left(\mathbb{C}^{4}\right)$.
It is known \cite{Kus2009} that every state $\ket{\psi}\in\mathrm{Fock}_{+}\left(\mathbb{C}^{4}\right)$
can be written, up to phase, as a combination
\begin{equation}
\ket{\psi}=\sqrt{1-a^{2}}\ket{\psi_{1}}+ia\ket{\psi_{2}}\,,\label{eq: real decomp}
\end{equation}
 where $0\le a\le\frac{1}{\sqrt{2}}$, and $\ket{\psi_{1}},\,\ket{\psi_{2}}$
are orthogonal states satisfying $\theta_{+}\ket{\psi_{\alpha}}=\ket{\psi_{\alpha}}$,
$\alpha=1,2$ (the latter condition corresponds to $\kb{\psi_{\alpha}}{\psi_{\alpha}}$
having real coefficients in the decomposition (\ref{eq:irrep decomp})).
Furthermore, by \cite{detection2012}, property (\ref{eq:irrep decomp})
implies that $\mathrm{Spin}(8)$ acts transitively on orthogonal pairs
$\ket{\psi_{1}},\ket{\psi_{2}}$ as above, whence we conclude that
the set of all states $\ket{\psi}$ corresponding to a given value
of $a$ is an orbit. In particular, $a=\frac{1}{\sqrt{2}}$ describes
Gaussian states, while $a=0$ is the orbit of $\ket{a_{8}}$ %
\footnote{For intermediate values of $a$, the presentation in terms of $\ket{\psi_{1}},\,\ket{\psi_{2}}$
is unique up to an overall sign; for $a=\frac{1}{\sqrt{2}}$ it is
unique up to rotation in the plane spanned by $\ket{\psi_{1}},\,\ket{\psi_{2}}$;
for $a=0$ only $\ket{\psi_{1}}$ is relevant, and is unique up to
sign. Accordingly the real dimension of the orbit is 7 for $a=0$,
$12$ for $a=\frac{1}{\sqrt{2}}$ and 13 for intermediate values of
$a$.%
}. Note that $C_{+}(\ket{\psi})=1-2a^{2}$ , so that $a$ is uniquely
determined by $C_{+}(\ket{\psi})$. It is also possible to re-express
the state $\ket{\psi}$, up to a phase, as a combination of orthogonal
Gaussian states (a generalised Schmidt decomposition). Namely, setting
$\ket{\psi_{G}}=\frac{1}{\sqrt{2}}\left(\ket{\psi_{1}}-i\ket{\psi_{2}}\right)$,
we have
\begin{equation}
\ket{\psi}=\sqrt{1-p^{2}}\ket{\psi_{G}}+p\theta_{+}\ket{\psi_{G}}\,,\label{eq:generalised Schmidt}
\end{equation}
where $p=\frac{1}{\sqrt{2}}\left(\sqrt{1-a^{2}}-a\right)$. Gaussianity
and orthogonality of $\ket{\psi_{G}},\,\theta_{+}\ket{\psi_{G}}$
is immediately verified. It is important to remark that in (\ref{eq:generalised Schmidt})
the phase of $\ket{\psi_{G}}$ \emph{does} matter (unless $\ket{\psi}$
is itself Gaussian).

The proof of (\ref{eq:Fidelity}) relies on Theorem 2 of \cite{Streltsov2010}
which, for our purposes, states that $F_{\mathrm{Gauss}}\left(\rho\right)$
can be described as a convex roof extension of a function defined
on pure states:
\begin{equation}
F_{\mathrm{Gauss}}\left(\rho\right)=\mathrm{sup}_{\sum p_{i}\kb{\psi_{i}}{\psi_{i}}=\rho}\left(\sum_{i}p_{i}F_{\mathrm{Gauss}}\left(\ket{\psi_{i}}\right)\right)\,,\label{eq:fidedity decomp}
\end{equation}
where $F_{\mathrm{Gauss}}\left(\ket{\psi}\right)=\mathrm{max}_{\kb{\phi}{\phi}\in\mathcal{G}}\, F\left(\kb{\psi}{\psi},\,\kb{\phi}{\phi}\right)$.
Using the fact that for $\ket{\psi}\in\mathrm{Fock}_{+}\left(\mathbb{C}^{4}\right)$
we have the decomposition (\ref{eq:generalised Schmidt}) we find
that $F_{\mathrm{Gauss}}\left(\ket{\psi}\right)=\tilde{F}\left(C_{+}\left(\ket{\psi}\right)\right),$
where $\tilde{F}(x)=\frac{1}{2}+\frac{1}{2}\sqrt{1-x^{2}}$ is a strictly
concave decreasing function on the interval $\left[0,1\right]$. Let
$\rho=\sum_{i}p_{i}\kb{\psi_{i}}{\psi_{i}}$ be the optimal decomposition
of $\rho$ leading to (\ref{eq:gen concurrence}). From \cite{Uhlmann1999}
it follows the all pure states in this decomposition have the same
value of the generalised concurrence i.e. $C_{+}\left(\ket{\psi_{i}}\right)=C_{+}\left(\rho\right)$.
Using this fact and (\ref{eq:fidedity decomp}) we have $F_{\mathrm{Gauss}}\left(\rho\right)\geq\tilde{F}\left(C_{+}\left(\rho\right)\right)$.
On the other hand, by concavity of $\tilde{F}$ we have
\[
\sum_{i}p_{i}F_{\mathrm{Gauss}}\left(\ket{\psi_{i}}\right)\leq\tilde{F}\left(\sum_{i}p_{i}C_{+}\left(\ket{\psi_{i}}\right)\right)=\tilde{F}\left(C_{+}\left(\rho\right)\right)\,,
\]
which concludes the proof of (\ref{eq:Fidelity}).

Summarising, we have presented a complete analytical characterisation
(\ref{eq:criterion}) of convex-Gaussian states in the four-mode fermionic
Fock space $\mathrm{Fock}_{+}\left(\mathbb{C}^{4}\right)$. Using
methods taken from entanglement theory and theory of Lie groups we
described quantitatively (see (\ref{eq:Fidelity}) and (\ref{eq:statistical distance}))
how the property of being non convex-Gaussian is resilient to noise.
These results have immediate consequence for the computation power
of models in which FLO or TQC with Ising anyons are assisted with
a noisy ancilla state. This follows form the fact that computations
augmented with convex-Gaussian states are classically simulable. We
have used our methods to give a precise value of the noise threshold
$p_{cr}=\frac{8}{11}$ above which the state $\ket{a_{8}}$ (used
to make FLO or TQC computationally universal \cite{powernoisy2013}),
when depolarised, becomes convex-Gaussian. This result is especially
interesting as the threshold value $p_{cr}$ is much higher than previously
known lower bounds. This opens a possibility for the existence of
$\ket{a_{8}}$-distillation protocols based on FLO or TQC that are
much more noise tolerant than the currently known ones. It would be
also intriguing to explore, using the analogy with entanglement theory,
the possible resource theory based on FLO or TQC with Ising anyons.
However, preliminary studies show that a ``naive'' application of
the entanglement purification protocol \cite{Bennett1996} based on
the generalised Schmidt decomposition (\ref{eq:generalised Schmidt})
is not applicable. In the future we would also like to characterise
the set of convex-Gaussian states defined on $\mathrm{Fock}\left(\mathbb{C}^{m}\right)$,
for $m>4$. To tackle this problem we plan to develop a theory of
non-linear witnesses introduced in \cite{Oszmaniec2013}.
\begin{acknowledgments}
We would like to thank Piotr \'Cwikli\'nski, Ravindra Witold Chhajlany
and Remigiusz Augusiak for fruitful discussions. The support of the
ERC grant QOLAPS is gratefully acknowledged. MO and MK acknowledge
the support of Polish National Science Centre grant under the contract
number DEC-2011/01/M/ST2/00379. MK acknowledges the support of COST
Action MP 1006.
\end{acknowledgments}

\bibliography{gausslit}

\end{document}